# Enhanced spontaneous emission in photonic hypercrystals

T. Galfsky[1,2], E.E. Narimanov[3], and V. M. Menon[1,2,*]

[1] Department of Physics, City College, City University of New York (CUNY), New York, USA
[2] Department of Physics, Graduate Center, City University of New York (CUNY), New York, USA
[3] Birck Nanotechnology Center, School of Computer and Electrical Engineering, Purdue University, Indiana, USA
*corresponding author: vmenon@ccny.cuny.edu

**Abstract:** We demonstrate a two-dimensional photonic hypercrystal that shows enhanced spontaneous emission from its metamaterial component and light extraction through its photonic crystal property. Spontaneous decay rate is enhanced by a factor of 19.5 and light extraction from the HMM is enhanced by a factor of ~100.

## 1. Introduction

Photonic crystals and metamaterials are two of the major paradigms in the field of photonics that have resulted in a plethora of discoveries both of fundamental and technological importance. These engineered optical materials rely on different physical mechanisms for their operation. Photonic crystals build upon Bragg scattering due to the periodicity in the dielectric environment where as metamaterials rely on average response of their sub-wavelength sized substructures (meta-atoms). The former implies that the feature sizes are on the same length scale as the wavelength of light being manipulated. In contrast, metamaterials require sub-wavelength feature sizes achieve an effective medium response which is the property that allows control of the flow of light in such materials. The combination of the two paradigms promises a new level of control in light propagation and enhancement. Among the different types of metamaterials, hyperbolic metamaterials (HMM), so called due to the hyperbolic shape of their iso-frequency surface, have received much attention in the recent times. This is in part due to their ease of fabrication, broadband response, leading to super resolution imaging and enhanced Purcell effect [1–6]. The unique optical properties of HMMs arise from the fact that they can support very large wavevector modes. Hence introducing any kind of periodicity in a HMM allows one to realize the photonic crystal features such as band edge modes and efficient scattering. The resulting photonic hypercrystals (PHC) are expected to show properties of both photonic crystals and HMMs [7]. While self-assembly of ferrofluids was used to demonstrate the formation of a PHC in the near infrared frequency range [8], so far there has been no demonstration of a hypercrystal for visible light. Here we report the first demonstration of an active PHC that shows both enhanced spontaneous emission stemming from its HMM component and enhanced light scattering to the far field due to its photonic crystal nature – thus resulting in a hypercrystal based light emitter. We show a decay rate enhancement factor of 19.5 as well as enhanced light extraction to far field by a factor of 66 over a spectral range of ~ 40 nm from colloidal quantum dots embedded inside the structure.

## 2. HMM design for broadband enhancement of spontaneous emission

Our HMM is composed of seven alternating periods (7P) of aluminum-oxide ($Al_2O_3$ ~ 16nm) and silver (Ag ~9nm) as seen in the schematic in Fig. 1a. The portrayed structure has an effective permittivity which is anisotropic in the *x-y* plane versus *z* direction. The equation for the isofrequency contour (IFC) for this kind of anisotropy is given by:

$$\frac{k_x^2 + k_y^2}{\varepsilon_z} + \frac{k_z^2}{\varepsilon_{x,y}} = \frac{\omega^2}{c^2} \qquad (1)$$

Where $\varepsilon_z$ and $\varepsilon_{x,y}$ are the components of the permittivity tensor in the z and x-y plane respectively. $k_x$, $k_y$, $k_z$ are components of the wavevector. $\omega$ and $c$ are the angular frequency, and the speed of light in vacuum. The effective permittivity of structure is plotted in Fig. 1b and the transition wavelength into the hyperbolic regime, $\lambda = 426nm$, is marked by a dashed line. At this wavelength the real part of the parallel permittivity component, $\varepsilon_{x,y}$, becomes negative thus changing the shape of IFC from an ellipsoid to hyperboloid (see schematics).

This topological transition of the IFC from a closed-faced to open-faced contour leads to a strong increase in the local photonic density of states (LPDOS) [9,10], similar to the Lifshitz transition of the Fermi surface of metals [5]. When dipole emitter such as molecular dyes or quantum dots is in close proximity to an HMM this enhancement in the LPDOS is expressed as enhancement of the dipole's decay rate, Γ, also known as Purcell factor [9]:

$$P = \frac{\Gamma_{HMM}}{\Gamma_0} \qquad (2)$$

Where $\Gamma_0$ is the radiative decay rate of a dipole in homogenous medium and $\Gamma_{HMM}$ is the dipole decay rate modified by the hyperbolic medium.

Most work on the subject was executed with a dipoles layer or a single dipole emitter placed on top a HMM [1–5]. However by embedding the active layer inside the HMM a much stronger broadband enhancement can be achieved. Fig. 2 shows the wavelength resolved LPDOS mapped as a function of wavevector for a dipole placed on top (panel a) and inside (panel b) a HMM, and panel c graphs the wavelength resolved Purcell factor. Notice that for a top placed dipole the radiative enhancement drops to 1/5 of the initial peak at the transition wavelength. In contrast, the embedded dipole layer maintains broadband enhancement which is 5 fold stronger than the dipole-on-top scenario. Of course the improved decay rate enhancement comes about by the stronger confinement of the electric field inside the metamaterial which presents the problem of out-coupling the dipole radiation into the far-field. Fig. 3 graphs a cross-section of the electric field generated by a dipole on top (panel a) and inside (panel b) a 7P HMM. The trapped radiation can be out-coupled to free space by nano-patterning as shown by our previous work [6] and others [11,12]. In section 4 we will show that by patterning the HMM into a PHC an out-coupling contrast of 100 fold over a spectral range of 40nm can be achieved.

### 3. Fabrication of a photonic hypercrystal (PHC)

We begin the fabrication with pre-cleaned glass micro-slides on which the active HMM (schematic shown in Fig. 4a) is fabricated by electron-beam deposition of seven alternating layers of Ag (~9nm) and $Al_2O_3$ (~16nm). The active layer of CdSe/ZnS quantum dots (QDs) (635nm center of emission) is embedded in the middle of the $5^{th}$ dielectric layer by spin coating. The multilayer is terminated with a 5nm thick capping layer of $Al_2O_3$. For every thin Ag layer a Ge seed layer of thickness ~1nm is deposited first. This ensures the formation of optically smooth Ag films which is highly important for propagation of surface plasmon-polaritons in the HMM [13,14]. Fig. 4b is a TEM micrograph of showing the continuous layers of Ag and $Al_2O_3$ (the Ge seed is indistinguishable in the image). In the next step of fabrication an array of nano-hole array patterned in the HMM by focused ion beam (FIB) (See SEM image in Fig 4c). The Holes are milled down to the QDs layer. By changing the hole radius, $r$, and lattice constant, $a$, an array of different photonic crystals was imprinted in the HMM, thus creating photonic hypercrystals (Fig. 5a). The effect of the PHCs on the QDs lifetime and emission properties is investigated in the following section.

### 4. Photoluminescence and lifetime of QDs in photonic hypercrystals

We use a home built confocal microscope with a scanning stage to observe the photoluminescence from the QDs. The QDs are excited by a pulsed diode laser at wavelength $\lambda = 440$nm with a repetition rate of 40MHz, emission is separated by a long pass filter and detected by an avalanche photodiode. The setup also allows to measure the emission lifetime at every pixel of the collected image. The technique is known as fluorescence-lifetime imaging microscopy (FLIM). Fig. 5b maps the photoluminescence (PL) from the array of PHCs shown in Fig. 5a where the dependence on the parameters $r$ and $a$ is evident. In addition to the PL map a lifetime map of the structures is shown in Fig. 5c. The out-coupling efficiency of the PHC is very clearly demonstrated by the sharp contrast of the PL emission from the patterned regions to the surrounding dark regions. Unlike a conventional photonic crystal which relies on resonance the enhancement effect of the PHC is broadband. Fig. 6a shows a wavelength resolved PL spectrum taken from the PHC with $r = 80$nm and $a = 280$nm. We notice that the enhancement effect spans the entire spectrum of the QDs with a contrast ratio of ~100 between PHC and unpatterned HMM. In Fig. 6b the lifetime kinetics curve of the same PHC is plotted together with the lifetime curves of an unpatterned HMM region and the curve for QDs on a glass substrate. Both lifetimes of the HMM and PHC are strongly modified from that of QDs on glass by factors of 9 and 19.5 respectively. We notice that the lifetime in the PHC region is much shorter than in the non-patterned region. This stems from the fact that the out-coupled light is preferentially composed of emission from the vertical components of the dipole ensemble which have better coupling the TM polarized high-k modes and therefore experience larger enhancement and are also better out-coupled to the far-field.

## 5. Discussion and conclusion

In order to investigate the effect of nano-holes another array was written with a very large lattice constant, $a = 600$nm. At this distance the coupling between nano-holes becomes weak and each hole can be resolved by out confocal microscope. Fig. 7a maps the PL intensity from this stretched PHC. We observe that each nano-hole acts as an independent scatterer of high-k modes. The out-coupling efficiency of this PHC is an order of magnitude lower than the maximal out-coupling preciously observed (Fig. 7b). The reduced efficiency as the separation between scattering centers is increasing comes about due to reduced coupling between scattering centers and plasmonic loss as the electromagnetic waves propagate a longer distance.

In conclusion, we present the first demonstration of an active PHC composed of a nano-patterned HMM. The device is shown to enhance light scattering from embedded emitters by a factor of ~100 while maintaining a Purcell factor of 19.5. This demonstration of a light emitting PHC is a step towards miniaturized photonic devices such as sub-wavelength lasers, ultrafast LEDs, and broadband single photon emitters.

**Acknowledgments**

Army Research Office (ARO) (W911NF15-1-0019); Army Research Office - MURI; NSF DMR MRSEC program – 1120923; Gordon and Betty Moore Foundation. Research was carried out in part at the Center for Functional Nanomaterials, Brookhaven National Laboratory, which is supported by the U.S. Department of Energy, Office of Basic Energy Sciences, under Contract No. DE-AC02-98CH10886.

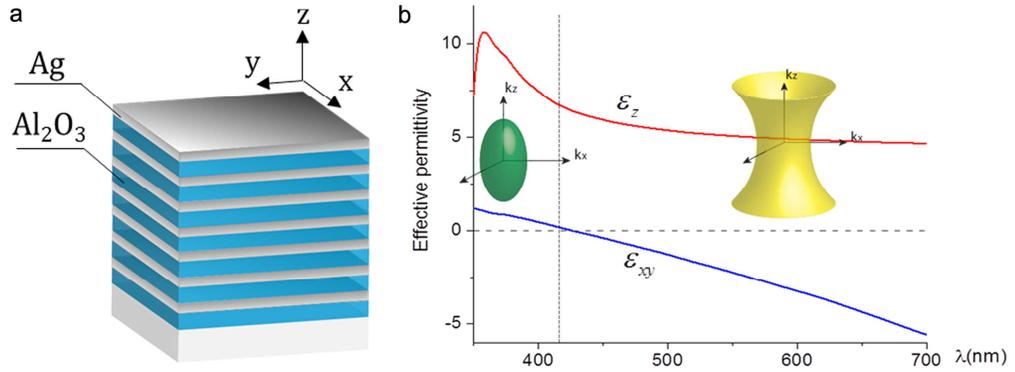

**Figure 1.** (a) Schematic of HMM composed of 7 periods (7P) of $Al_2O_3$ and Ag (b) Effective permittivity tensor components. Dashed line at 426nm marks the transition wavelength from elliptical to hyperbolic regime.

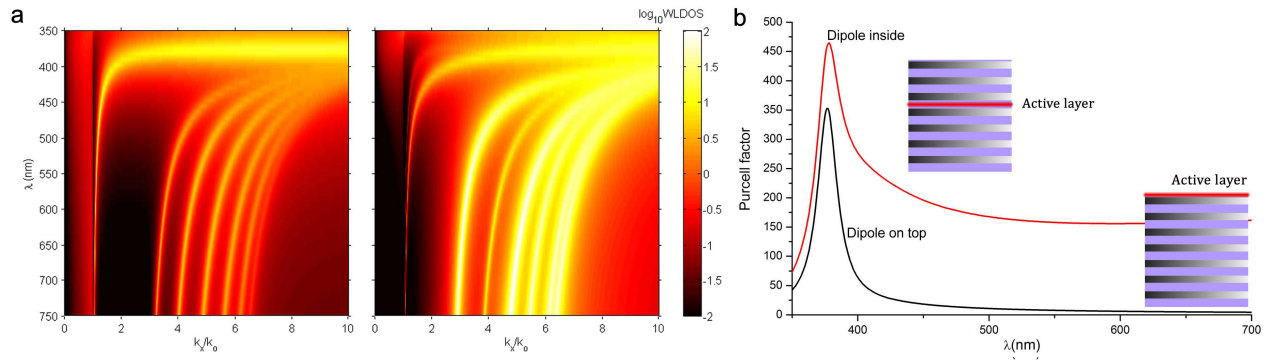

**Figure 2.** (a) Wavelength resolved local photonic density of states for dipole on top 7P (left) and embedded inside (right) 7P HMM. (b) Purcell factors as a function of wavelength for the two cases with schematics.

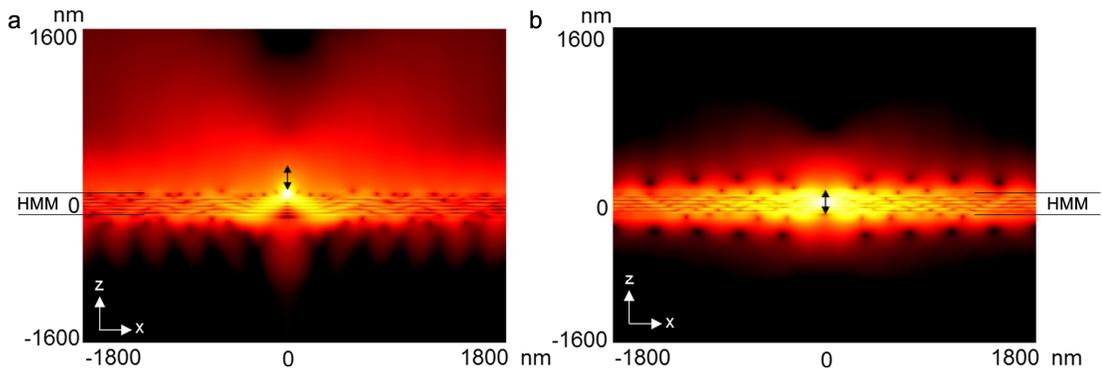

**Figure 3.** Simulation of the electric field generated by an electric dipole. (a) Placed on top of the HMM, (b) placed inside the HMM, demonstrating stronger confinement of the electric field.

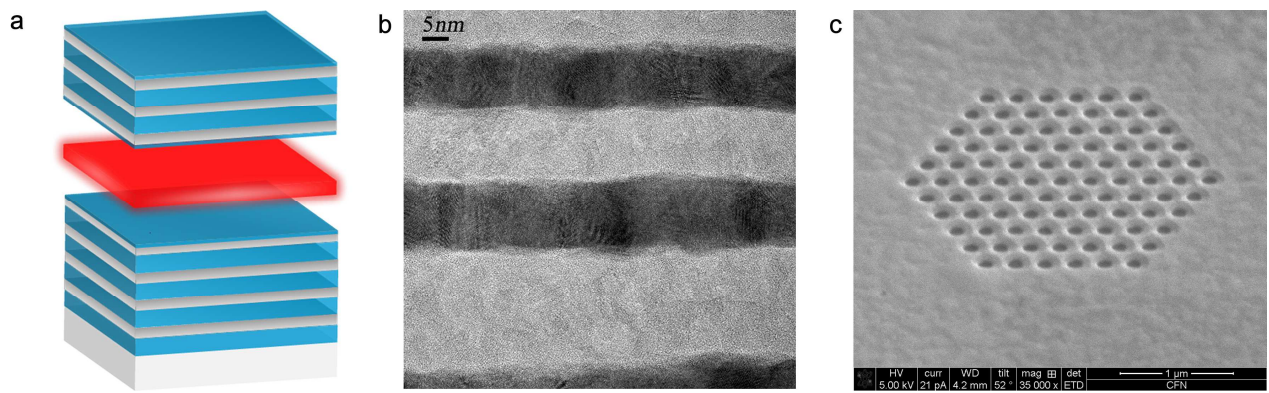

**Figure 4.** (a) Schematic of active hyperbolic metamaterial (b) TEM micrograph of Ag (dark) and $Al_2O_3$ (light) layers in the fabricated HMM (c) SEM micrograph of a photonic hypercrystal composed of an array of nano-holes patterned in the HMM with focused ion beam milling.

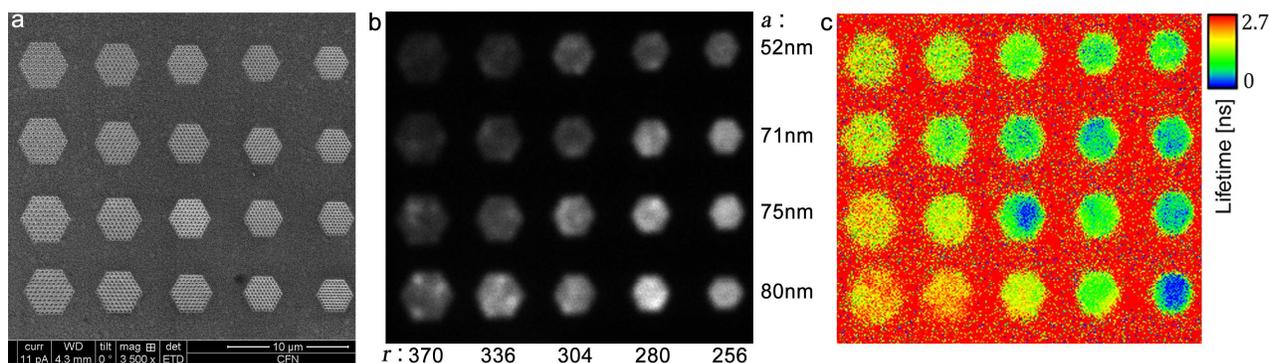

**Figure 5.** (a) SEM image of patterned PHCs. (b) Photoluminescence intensity map of array. (c) Lifetime map of the patterned array.

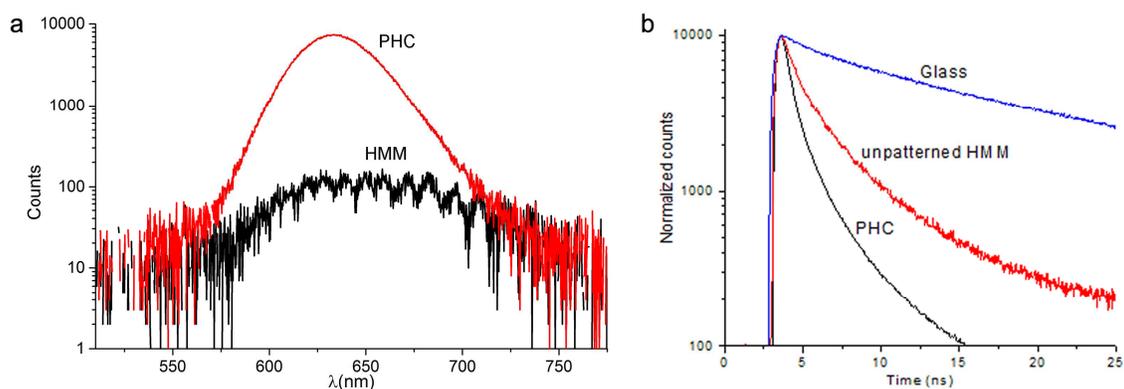

**Figure 6.** (a) Photoluminescence spectrum of PHC and HMM. The observed wiggles in the HMM curve are due to interference effect from the emission filter in the optical path (b) Lifetime kinetics of QDs embedded in, HMM, PHC and QDs on a glass substrate. The respective lifetimes are 2.97ns, 1.28ns, and 25ns.

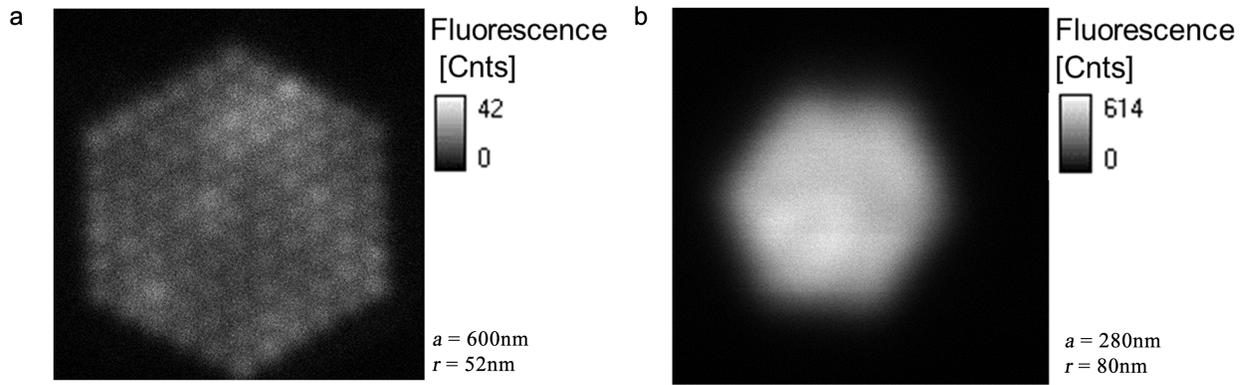

**Figure 7.** (a) Photoluminescence image of PHC with a=600nm and r=52nm, each nano-hole can be seen to act as an independent scatterer (b) Photoluminescence image of PHC with a=280nm and r=80nm, coupling between nano-holes leads to improved out-coupling efficiency.